\documentclass[11pt,a4paper]{article}
\pdfoutput=1

\usepackage{Packages}

\usepackage{Definitions}

\allowdisplaybreaks[1]

\newcommand{\OfficialTitle}{The Gauge-Bethe Correspondence and Geometric Representation Theory}

\hypersetup{pdfauthor={Domenico Orlando and Susanne Reffert},pdftitle={\OfficialTitle}}

\author{
  \begin{minipage}{.97\linewidth}
    \vspace{1cm}
    \begin{center}
      {\small
        \textbf{Domenico Orlando} and \textbf{Susanne Reffert}
      }
    \end{center}
    \vspace{1cm}\begin{minipage}{\linewidth}
      {\itshape \footnotesize \centering
        Institute for the Mathematics and Physics of
        the Universe, \\The University of Tokyo, Kashiwa-no-Ha
        5-1-5, \\ Kashiwa-shi, 277-8568 Chiba, Japan.\\
      }
    \end{minipage}
    \vspace{1cm}
    \end{minipage}
}

\date{}

\title{\vspace{2cm}
  {\huge   \textbf{\OfficialTitle}}
}

\begin{document}

\numberwithin{equation}{section}

\begin{titlepage}
  \maketitle
  \thispagestyle{empty}

  \vspace{-14cm}
  \begin{flushright}
    IPMU10-0217
  \end{flushright}

  \vspace{18cm}

  \begin{center}
    \textsc{Abstract}
  \end{center}
  The Gauge/Bethe correspondence of Nekrasov and Shatashvili relates
  the spectrum of integrable spin chains to the ground states of
  supersymmetric gauge theories. Up to now, this correspondence has
  been an observation; the underlying reason for its existence
  remaining elusive.  We argue here that geometrical representation
  theory is the mathematical foundation of %
  the Gauge/Bethe correspondence, and it provides a
  framework to study families of gauge theories in a unified way.

\end{titlepage}

\clearpage
\section{Introduction}
\label{sec:introduction}

In~\cite{Nekrasov:2009uh, Nekrasov:2009ui} the \emph{Gauge/Bethe correspondence} was introduced, relating integrable spin chains to 
$2d$ supersymmetric gauge theories. The duality was 
further extended to four dimensional gauge theories in~\cite{Nekrasov:2009rc}. In~\cite{Orlando:2010uu}, it was used to relate the ground 
states of different quiver gauge theories. 
In the simplest case of a $\mathrm{su}(2)$ \textsc{xxx}${}_{1/2}$ spin chain, the statement of this correspondence can be summarized as follows: the spectrum of the $N$ magnon sector of the integrable system 
corresponds to the supersymmetric ground states of a matching $\mathcal N=(2,2)$ $\mathrm{U}(N)$ gauge theory in two dimensions. For an \textsc{xxx}${}_{1/2}$ spin chain of length $L$, the 
low energy limit of the corresponding gauge theory is the non-linear sigma model (\textsc{nlsm}) with target space the cotangent bundle of the Grassmannian $T^*\gr
(N,L)$~\cite{Nekrasov:2009ui}. 
\bigskip

Since the total spin is a symmetry of the spin chain, one usually decomposes the 
Hilbert space of the spin chain into $N$ \emph{magnon} sectors, \emph{i.e.} sectors with $N$ down spins and total spin $L/2-N$. To each such sector 
with $N$ magnons corresponds a $\mathrm{U}(N)$ gauge theory  via the Gauge/Bethe construction. 
To describe the full Hilbert space and the full spectrum of the spin
chain, of course all $N$ magnon sectors with $N = 0,\dots,L$ need to
be considered.  In gauge theory, on the other hand, one usually
studies each $\mathrm{U}(N)$ by itself, and not different values of $N$
together. There is, however, evidence that doing so would make
sense. In the case of the \textsc{xxx}${}_{1/2}$ spin chain, there is
an obvious equivalence between taking the state with all spins up as
the reference state and considering the sector with $N$ down spins,
and taking the state with all spins down as reference state and
considering the sector with $L-N$ up spins instead. This equivalence
is on the gauge theory side reflected in the \emph{Grassmannian
  duality} relating $\gr(N,L)$ and $\gr(L-N,L)$. This is the simplest
instance of a relation between gauge theories with different $N$.

We would like to argue that taking the Gauge/Bethe correspondence seriously, one should look for a framework in which it makes sense to consider gauge theories with 
different $N$ together. Moreover, we learn from the integrable model side of the correspondence that this set of gauge theories must carry a $\mathrm{su}(2)$ 
symmetry, a fact that only becomes obvious in such a unified framework. The integrable structure of the spin chain remains hidden as long as the gauge theories are considered separately, but will be manifest in a framework that unifies them in a meaningful way.
In fact, such a mathematical framework exists and goes under the name of \emph{geometric representation theory}. It goes back to Victor 
Ginzburg and is explained in detail in his book~\cite{Chriss:1997}.
Up to now, the Gauge/Bethe correspondence has been an observation; the
underlying reason for its existence remaining elusive.  Geometrical
representation theory can be understood as the mathematical foundation
or underlying reason for the Gauge/Bethe correspondence.

\bigskip
We will be explicitly treating the $\mathrm{su}(2)$ Heisenberg spin chain, which is the simplest case. 
However, our arguments extend directly to $\mathrm{su}(n)$ spin chains for any $n\in \mathbb{N}$. The corresponding sigma models in this case have 
as target space the cotangent bundle of a \emph{flag variety}.
While the Gauge/Bethe correspondence is applicable to spin chains with \emph{any} Lie group or even supergroup symmetry, the 
mathematical construction of the geometric representation theory for the general case is beyond the scope of this note.

\bigskip
The plan of this note is as follows. In Sec.~\ref{sec:gauge-bethe}, we briefly summarize the %
the Gauge/Bethe 
correspondence, leaving the details to the original literature. In Sec.~\ref{sec:geom-repr-theory}, 
geometric representation theory is introduced and applied to the Gauge/Bethe correspondence. In Sec.~\ref{sec:further-directions}, the 
implications of the above and further directions of study are explored.

\section{The Gauge/Bethe correspondence}
\label{sec:gauge-bethe}

The Gauge/Bethe correspondence, as detailed in~\cite{Nekrasov:2009uh,
  Nekrasov:2009ui}, relates two-dimensional $\mathcal{N}=(2,2)$
supersymmetric gauge theories to quantum integrable systems. The
supersymmetric vacua of the gauge theories form a representation of
the \emph{chiral ring}, which is a distinguished class of operators
which are annihilated by one chirality of the supercharges $Q$. The
\emph{commuting Hamiltonians} of the quantum integrable system are
identified with the generators of the chiral ring. The space of states
of the quantum integrable system, \emph{i.e.} the spectrum of the
commuting Hamiltonians, is thus mapped to the supersymmetric vacua of
the gauge theory.
Expressed differently, the \emph{effective twisted superpotential} of
the gauge theory in the Coulomb branch is identified with the
\emph{Yang--Yang counting function} which serves as a potential for
the Bethe equations, whose solutions are the spectrum of the
integrable system. This is true sector by sector.
The dictionary between a general spin chain and the corresponding gauge theory was detailed in~\cite{Orlando:2010uu}. 
Here, we will 
confine ourselves to a very simple case, namely the one of a $\mathrm{su}(n)$ (mostly $n=2$) Heisenberg spin chain of length $L$ with periodic 
boundary conditions and no inhomogeneities, where each position carries the fundamental representation of $\mathrm{su}(n)$. Each position in the 
chain admits one of $n-1$ different particle species. If we focus on the sector containing $N_a$ particles of species $a$, this model results in 
an $\mathrm{A}_{n-1}$ quiver gauge theory with the following properties: each node $a=1,\dots,r$  carries a $\mathrm{U}(N_a),\ N_a\in\{1,\dots,L\}$ gauge 
group. The bifundamental matter fields between neighboring nodes have twisted mass $m_B=-\imath/2$, the adjoint matter fields occurring at each 
node have twisted mass $m_\Phi=\imath$. One of the $n-1$ nodes is connected to a $\mathrm{U}(L)$ flavor group, and the fundamental and anti-fundamental 
matter field have twisted mass $m_Q = - \imath/2$.

The simplest case, namely that of the $\mathrm{su}(2)$ \textsc{xxx} spin chain of
length $L$, corresponds to the $\mathcal{N}=(2,2)$ theory with gauge group $\mathrm{U}(N)$, flavor group $\mathrm{U}(L)$ and
the following matter content:
\begin{itemize}
\item an adjoint field $\Phi$ with twisted mass $\imath$,
\item $L$ fundamentals and anti--fundamentals $Q_n, \overline{Q}_n$
  with twisted mass $-\imath /2 $.
\end{itemize}
We are interested in the \emph{Coulomb branch} of the theory: we
therefore consider the low energy effective theory obtained for slowly
varying $\sigma$ fields\footnote{The complex scalar $\sigma$ is the lowest component of the super-field
strength, which is a twisted chiral multiplet.}, after integrating out the massive matter fields. The
resulting vacuum manifold is the cotangent bundle to the Grassmannian
$T^* \gr(N,L)$:
\begin{align}
  \gr(N,L) &= \set{W\subset \setC^L | \dim W = N} \, , \\
  T^* \gr(N,L) &= \set{(X,W), W \in \gr(N,L), X \in
    \mathrm{End}(\setC^L) | X(\setC^L) \subset W,
    X(W) = 0} \, .
\end{align}
Via the effective twisted superpotential
\begin{multline}
 \widetilde W^N_{\text{eff}} ( \sigma ) = \frac{L}{2\pi} \sum_{i=1}^{N}
  \left[ \left( \sigma_i + \frac{\imath}{2} \right) \left( \log
      (\sigma_i + \frac{\imath}{2}) - 1 \right) - \left( \sigma_i -
      \frac{\imath}{2} \right) \left( \log (-
      \sigma_i + \frac{\imath}{2}) - 1 \right) \right] \\
  + \frac{1}{2\pi} \sum_{\substack{i,j\\i \neq j }}^{N} \left(
    \sigma_i - \sigma_j - \imath \right) \left( \log (\sigma_i -
    \sigma_j - \imath ) - 1 \right) - \imath \tau \sum_{i=1}^{N}
  \sigma_i \,,
\end{multline}
we can obtain the vacua of the theory which are the solutions of the
equation~\cite{Hori:2003ic}
\begin{equation}
 \exp \left[ 2\pi \frac{\partial \widetilde W_{\text{eff}}(\sigma)}{\partial \sigma_i} \right]  = 1 \, .
\end{equation}
The minima satisfy the equation
\begin{equation}
\label{eq:gauge-minima}
  \left( \frac{\sigma_i + \imath /2}{\sigma_i - \imath/2} \right)^L
  = \prod_{\substack{ j = 1 \\ j \neq i}}^N \frac{\sigma_i -
    \sigma_j + \imath}{\sigma_i - \sigma_j - \imath} \hspace{2em} i
  = 1,2, \dots, N \, .
\end{equation}
Because of the presence of twisted masses, the corresponding ground states
generate the equivariant cohomology of $H^*_{\mathrm{SU}(L) \times
  \mathrm{U}(1)}[T^*\gr(N,L)]$ which in this case (\emph{i.e.} no
inhomogeneities in the spin chain) is isomorphic to the standard
cohomology.

\bigskip

The other side of the correspondence is the \textsc{xxx}${}_{1/2}$ spin chain,
\emph{i.e.} a system of $L$ spins on a circle, each of which carries
the fundamental representation $V= \setC^2$ of $\mathrm{su}(2)$. The total
Hilbert space is given by the product
\begin{equation}
  \label{eq:product-xxx}
  \mathcal{H} = \bigotimes_{n=1}^L V \simeq \left( \setC^2
  \right)^{\otimes L} \, ,
\end{equation}
and there is a natural action of $\mathrm{su}(2)$ on $\mathcal{H}$, generated by
\begin{align}
  E &=\sum_{m=1}^L \underbrace{\mathbbm{1} \otimes \dots \otimes \mathbbm{1}
  }_{m-1}\otimes\, e \otimes\mathbbm{1}\otimes\dots\otimes\mathbbm{1}\,, \\
  F &=\sum_{m=1}^L \underbrace{\mathbbm{1} \otimes \dots \otimes \mathbbm{1}
  }_{m-1}\otimes\, f \otimes\mathbbm{1}\otimes\dots\otimes\mathbbm{1}\,, \\
  K &=\sum_{m=1}^L \underbrace{\mathbbm{1} \otimes \dots \otimes \mathbbm{1}
  }_{m-1}\otimes\, k \otimes\mathbbm{1}\otimes\dots\otimes\mathbbm{1}
  \, , 
\end{align}
where $e,f,k$ generate the fundamental representation of $\mathrm{su}(2)$:
\begin{align}
  e &=
  \begin{pmatrix}
    0 & 1 \\ 0 & 0
  \end{pmatrix}\,, &
  f &=
  \begin{pmatrix}
    0 & 0 \\ 1 & 0
  \end{pmatrix}\,, &
  k &=
  \begin{pmatrix}
    1 & 0 \\ 0 & -1
  \end{pmatrix}\,.
\end{align}
Physically, $E$ flips a spin up, $F$ flips a spin down and $K$ is
twice the total spin.
Thanks to the $\mathrm{su}(2)$ invariance of the Hamiltonian, the Hilbert space
decomposes into the direct sum of subspaces with fixed number of spins
up:
\begin{align}
  \mathcal{H} &= \bigoplus_{N=0}^L V_{L-2N}\,, \\
  V_{L-2N} &= \set{\psi \in \mathcal{H}| K\, \psi = \left( L - 2N
    \right) \psi } \, .
\end{align}
The vectors $\psi \in V_{L-2N}$ contain $N$ excitations, or magnons. A
basis of the weight space $V_{L-2N}$ is obtained as the set of states
depending on $N$ rapidities $\lambda_i$ that satisfy the Bethe Ansatz
equation\footnote{In the case of twisted boundary conditions, the
  Bethe Ansatz equation admits $\binom{L}{N}$ solutions. For periodic
  b.c. there are only $\binom{L}{N}- \binom{L}{N-1}$ solutions,
  corresponding to the highest weight states of $\mathrm{su}(2)$ in $V_{L-2N}$; these
  have to be supplemented by the descendants of the states in $V_{L-2N
    +2}$. For details see \emph{e.g.}~\cite{Faddeev:1996iy}.\\ This is
  consistent with the fact that the geometric interpretation of the
  gauge theory requires a finite \textsc{fi} term.}:
\begin{equation}
\label{eq:decomposition-xxx}
  \left( \frac{\lambda_i + \imath /2}{\lambda_i - \imath/2} \right)^L
  = \prod_{\substack{ j = 1 \\ j \neq i}}^N \frac{\lambda_i -
    \lambda_j + \imath}{\lambda_i - \lambda_j - \imath}\,, \hspace{2em} i
  = 1,2, \dots, N \, .
\end{equation}
This is precisely the same equation that we had found above for the
ground states of the non-linear sigma model once we identify the field
$\sigma_i$ with the rapidity $\lambda_i$. This observation is the
essence of the Gauge/Bethe correspondence.  Note that the solutions to
the Bethe Ansatz equation, which we find on both sides of the
correspondence, parametrize the \emph{eigenvectors} of the
\textsc{xxx}${}_{1/2}$ chain and not the eigenvalues of the Hamiltonian. 
We would like to stress that the properties of an integrable spin
chain are  determined by its symmetry group (including the information
about its boundary conditions and the representation of the symmetry
group each position is carrying). Its Hamiltonian is a consequence of the Yang--Baxter equation which is determined by the symmetries of the chain.

Given the Gauge/Bethe correspondence we can use techniques from
representation theory (the \textsc{xxx} side) to answer gauge
theory questions. The completeness of the Bethe equation for example
implies that the generating function for the number of solutions in
each sector $V_{L-2N}$ is given by the character of $q^{K}$. The latter
decomposes into the product over the $L$ fundamental representations:
\begin{equation}
  Z(q) = \Tr_{\mathcal{H}} q^{K} = \prod_{n=1}^L \Tr_{V} q^{k}
  = q^{L} \prod_{n=1}^L \left( 1 + q^{-2} \right) =
  \sum_{N=0}^L \binom{L}{N} q^{L-2N} \, .
\end{equation}
Via the correspondence, this is also the generating function for the
dimensions of the cohomologies of the cotangent bundles to the
Grassmanians. Hence
\begin{equation}
  \dim H^*[ T^* \gr(N,L) ]= \binom{L}{N} \, . 
\end{equation}

\vfill 

\section{Geometric representation theory for \textrm{su}(2)}
\label{sec:geom-repr-theory}

We now introduce the main ideas of geometric representation
theory. Precise details can be found in the book~\cite{Chriss:1997}.

A representation of the algebra $\mathrm{su}(2)$ consists in a vector
space $V$ and an action of three operators $e,f,k$ satisfying the relations
\begin{align}
  [e, f] &= k \,, & [k,e] &= 2 e \,, & [k, f] = -2f \, .
\end{align}
Given the tensor product of $L$ copies of the fundamental
representation $V$, there is a natural inclusion of the
$(L+1)$--dimensional irreducible representation, $V(L) \hookrightarrow
V^{\otimes L}$. Geometric representation theory provides a
construction for $V^{\otimes L}$ where the vector space is the direct
sum of the homologies of the cotangent bundles to all the
Grassmannians $T^* \gr(N,L)$ for fixed $L$:
\begin{equation}
  \label{eq:VL-grassmannians}
  V^{\otimes L} \simeq \bigoplus_{N=0}^L H_* [T^* \gr(N,L), \setC] = H_* [T^* \gr(L), \setC] \, , 
\end{equation}
where $ \gr(L)$ is the disjoint union of all the Grassmannians $ \gr(N,L)$
for fixed $L$:
\begin{equation}
   \gr(L) = \bigsqcup_{N=0}^L  \gr(N,L) 
\end{equation}
(we set $ \gr(0,L) = \gr(L,L) =\emptyset$).  Each of the terms $H_*
[T^* \gr(N,L), \setC]$ is identified with the $(L - 2N)$ weight space,
which has dimension $\binom{L}{N}$:
\begin{equation}
  H_*[T^* \gr(L)] \simeq V^{\otimes L} = \bigoplus_{N=0}^L V_{L-2N}
  \simeq \bigoplus_{N=0}^L H_* [T^* \gr(N,L)] \, . 
\end{equation}

The key point of the construction is the definition of the operators
$e$ and $f$ which act between the homologies,
\begin{equation}
  e,f : H_*[T^* \gr(L)] \to H_*[T^* \gr(L) ] \, .
\end{equation}
In particular, we need $f$ to act between two components of $ \gr(L)$,
raising $N$ by $1$:
\begin{equation}
  f : H_*[T^* \gr(N,L)] \to H_*[T^* \gr(N+1,L) ] \, .
\end{equation}
This is possible by introducing a \emph{correspondence} (in the mathematical sense)
\begin{equation}
  \begin{tikzpicture}[description/.style={fill=white,inner sep=2pt}]
    \matrix (m) [matrix of math nodes, row sep=3em, column sep=1em, text height=1.5ex, text depth=0.25ex]
    {  & Z \subset  T^*  \gr(N,L) \times T^*  \gr (N+1,L)  &  \\
      T^* \gr(N,L) &  & T^* \gr(N + 1,L) \\ }; 
    \path[>=latex,->,font=\scriptsize] 
    (m-1-2) edge node[auto] {$ {\pi_1} $} (m-2-1)
    edge node[auto] {$ {\pi_2} $} (m-2-3); 
  \end{tikzpicture}
\end{equation}
where $Z $ is the diagonal part of the cotangent bundle of the product of  two Grassmannians:
\begin{equation}
  Z = \set{(X, U_N, U_{N+1}) | U_i \in  \gr(N_i,L), X \in \mathrm{End}(\setC^L), U \subset U^\prime, X(\setC^L) \subset U, X(U^\prime) = 0 } \, .
\end{equation}
We can now define the \emph{Hecke operator} $f$ by first acting with the
pullback ${\pi_1}^*$, then intersecting with the fundamental class
$[Z]$ and finally acting with the pushforward ${\pi_2}_*$:
\begin{align}
  f \colon H_* [T^* \gr(N,L)] &\to H_* [T^* \gr(N+1,L)]  \\
  x &\mapsto f(x) = {\pi_2}_* ( [Z] \cap {\pi_1}^*(x) )\,.
\end{align}
The operator $e$ is defined in a similar way,
\begin{align}
  e \colon H_* [T^* \gr(N+1,L)] &\to H_* [T^* \gr(N,L)]  \\
  x &\mapsto e(x) = (-1)^L {\pi_1}_* ( [Z] \cap {\pi_2}^*(x) )\,.
\end{align}
One can prove that the commutation relation $[e,f]= k $ is satisfied
if $k$ is the operator that multiplies each component $H_* (T^*
\gr(N,L))$ by $(L - 2N)$:
\begin{equation}
  k = \bigoplus_{N=0}^L \left( L - 2N \right) \id_{H_* [T^*
\gr(N,L)]} \, ,
\end{equation}
which shows that the $L-2N$ weight space is precisely the homology of $T^* \gr(N,L)$:
\begin{equation}
  \label{eq:weight-space-Grassmann}
  H_* [ T^* \gr(N,L)] = \set{x \in H_* [ T^* \gr(L)] \simeq V^{\otimes L} | k x = \left( L - 2N \right) x} \simeq V_{L-2N} \, .
\end{equation}

In particular one finds that the homology $H_{top}[T^* \gr(L)]$ (the
span of the fundamental classes of all the Grassmannians of $\setC^L$,
$[T^*\gr(N,L)]$) is stable under the action of $e $ and $f$ and hence
provides the highest weight representation of dimension $L+1$,
\begin{equation}
  H_{top}[T^* \gr(L)] = \bigoplus_{N=0}^L  H_{top}[T^* \gr(N,L)]  = \bigoplus_{N=0}^L [T^* \gr(N,L)] \simeq V(L) \, .  
\end{equation}

Let us try to interpret this construction in terms of physical
systems. The direct sum of the homologies of the cotangent bundles
over the Grassmannians of $\setC^L$ corresponds to the ground states
of the non-linear sigma models on all the $T^* \gr(N,L)$ for $N=0,1,
\dots,L$.  Via geometric representation theory, this space can be
given the structure of the $V^{\otimes L}$ representation of $\mathrm{su}(2)$
(Eq.~\eqref{eq:VL-grassmannians}).  The Hilbert space of the
\textsc{xxx}${}_{1/2}$ spin chain (Eq.~(\ref{eq:product-xxx})) has the same
structure. We can identify the two spaces. The homology
of the Grassmannian $H_*[T^* \gr(N,L)]$ (Eq.~\eqref{eq:gauge-minima})
is the $(L-2N)$ weight space $V_{L-2N}$
(Eq.~\eqref{eq:weight-space-Grassmann}), which is spanned by the
spectrum of the \textsc{xxx}${}_{1/2}$ chain in the $N$ magnon sector
(Eq.~\eqref{eq:decomposition-xxx}).
It follows that there is a one-to-one
correspondence between the minima of the twisted superpotential of the
\textsc{nlsm} and the solutions to the Bethe Ansatz (gauge/Bethe
correspondence). In Table~\ref{tab:dictionary} we provide a minimal
dictionary between the gauge Bethe correspondence and geometric
representation theory.

The construction can be generalized to $\mathrm{A}_{n-1}$ algebras which are
represented by homologies of cotangent bundles to flag spaces $H_*
[T^* \mathrm{Fl}_n \setC^L]$:
\begin{align}
  \mathrm{Fl}_{\mu}\setC^L &= \set{ 0 = U_0 \subset U_1 \subset \dots \subset U_n = \setC^L, \mu \in \setN^n | \dim U_i - \dim U_{i-1} = \mu_i} 
\, , \\
  T^* \mathrm{Fl}_\mu \setC^L &= \set{(X,U_{\bullet})| U_{\bullet} \in \mathrm{Fl}_\mu \setC^L , X(U_i) \subset U_{i-1}} \, ,\\
  \mathrm{Fl}_n\setC^L &= \bigsqcup_{\mu_1+\mu_2+\dots+\mu_n = L} \mathrm{Fl}_\mu\setC^L \, .
\end{align}
These are the vacuum manifolds for the quiver gauge theories
identified in Sec.~\ref{sec:gauge-bethe}.

\begin{table}
  \centering
  \begin{tabular}{cc}
    \toprule
    physics & mathematics \\ \midrule
    spectrum of \textsc{xxx}${}_{1/2}$ spin chain & $\mathrm{su}(2)$ representation $V^{\otimes L} \simeq H_*[T^* \gr(L)]$ \\
    ground states of the \textsc{nlsm} on $T^*\gr(N,L)$ & cohomology $H^*[T^* \gr(N,L)]$ \\
    spectrum for the $N$ magnon sector & weight space $V_{L-2N} \simeq H_*[T^* \gr(N,L)]$\\
    ground states of \textsc{xxx}${}_{1/2}$ & hw representation $V(L) \simeq H_{top}[T^* \gr(L)]$ \\
    gauge/Bethe correspondence & geometric representation of $\mathrm{su}_2$ \\
    \bottomrule
  \end{tabular}
  \caption{Dictionary}
  \label{tab:dictionary}
\end{table}

\section{Further directions}
\label{sec:further-directions}

This note is a starting point for a wealth of generalizations and
further developments. Our statement readily generalizes to $su(n)$, but the scope of the Gauge/Bethe correspondence is much wider and includes spin chains with different symmetry groups, inhomogeneities, etc. While a number of mathematical difficulties need to be conquered in order to get there, we expect the formalism of geometric representation theory to be powerful enough to encompass all of the above.
We would like to conclude this note by pointing out some directions for
future research, based on the identification between the Gauge/Bethe
correspondence and geometric representation theory. 
\begin{enumerate}
\item The central object of the Gauge/Bethe correspondence is the
  Bethe Ansatz equation which is a consequence of the Yang--Baxter
  relations. These are in turn related to representations of the
  \emph{Yangian algebra} for which the spin chain admits a natural
  action~\cite{Faddeev:1996iy}.  This points to a representation of
  the Yangian in terms of Grassmannians, generalizing the construction
  of Sec.~\ref{sec:geom-repr-theory}, which would further elucidate
  the role of integrability.
\item The Gauge/Bethe correspondence associates products of general
  representations to \emph{equivariant} cohomologies. This calls for a
  generalization of the construction in
  Sec.~\ref{sec:geom-repr-theory} to the equivariant case.
\item Bethe Ansatz equations of the form~\eqref{eq:decomposition-xxx}
  can be written for quantum integrable systems with any Lie group, or
  supergroup symmetry. To capture the general case, geometric representations of all these groups are necessary.
\item The space of solutions of \textsc{xxx}${}_{1/2}$ admits a
  natural grading in terms of \emph{rigged partitions}~\cite{MR869577}. The corresponding
  fermionic formula reproduces the generating function for the
  Poincaré polynomials of the Grassmannians $\gr(L)$.
\end{enumerate}
Geometric representation theory is moreover closely related to Nakajima's
  work on quiver varieties~\cite{MR1604167,MR1865400}, which opens up further interesting avenues of study.
  \bigskip

Perhaps the most important message is that it makes sense to consider
a unified framework in which a set of gauge theories (\emph{e.g.} all
the $\mathrm{U}(N)$ gauge theories with $L$ flavors and $N=0,1, \dots,
L$) are studied together. Only in this setting, the $su(2)$
algebra (and in general the integrable structure) will become
manifest. In a string theory interpretation, the Hecke operators $e$
and $f$ defined in Sec.~\ref{sec:geom-repr-theory} could be naturally
understood as brane creation and annihilation operators. But this goes
beyond the scope of this note.

\subsection*{Acknowledgements}

It is a pleasure to thank Simeon Hellerman and Kentaro Hori
for inspiring discussions. We would furthermore like to thank Anthony Licata and Nicolai Reshetikhin for
correspondence.  This research
was supported by the World Premier
International Research Center Initiative (WPI Initiative), MEXT,
Japan.

\bibliography{GaugeBetheRef}

\end{document}